\pdfoutput=1
\documentclass[11pt]{article}
\usepackage[final]{acl}

\usepackage{times}
\usepackage{latexsym}

\usepackage[T1]{fontenc}
\usepackage[utf8]{inputenc}

\usepackage{microtype}

\usepackage{inconsolata}
\usepackage{subcaption}

\usepackage{float}

\usepackage{siunitx}
\usepackage{natbib}
\usepackage{amsmath}
\usepackage{amssymb}
\usepackage{booktabs}
\usepackage{adjustbox}
\usepackage{multirow}
\usepackage{geometry}
\usepackage{graphicx}
\usepackage{tcolorbox}

\usepackage{lipsum} % 예시용 텍스트

\usepackage[normalem]{ulem}
\usepackage{hyperref}
\usepackage{cleveref}% Load AFTER hyperref!
\usepackage{blindtext}
\usepackage[colorinlistoftodos]{todonotes}
\usepackage{caption}
\usepackage{amsmath}

\usepackage{authblk}

\usepackage{makecell}

\usepackage{twemojis}
\usepackage{wrapfig}

\usepackage{amssymb}% http://ctan.org/pkg/amssymb
\usepackage{pifont}% http://ctan.org/pkg/pifont

\usepackage{adjustbox}
\usepackage{graphicx}
\usepackage{tabularx}
\usepackage{parskip}
\usepackage{helvet} % DO NOT CHANGE THIS
\usepackage{courier}  % DO NOT CHANGE THIS
\usepackage{algorithm}
\usepackage{algpseudocode}
\usepackage{graphicx}
\usepackage{amsmath}

\usepackage[record,abbreviations]{glossaries-extra}
\glsdisablehyper

\usepackage[dvipsnames,table,xcdraw,svgnames]{xcolor}

\usepackage{minted}

\usepackage{placeins}

\newcommand{\mysymbol}{\texttwemoji{handshake: light skin tone, medium-dark skin tone}}

\newcommand{\ours}{UnIte\xspace}

\newcommand{\oursAbbr}{\textbf{Un}certainty-based \textbf{Ite}rative Document Sampling\xspace}

\newcommand{\oursFullHighlight}{\Large{\mysymbol}
\ours:
Uncertainty-based Iterative Document Sampling\\ for Domain Adaptation in Information Retrieval\xspace}

\newcommand{\oursTitle}{\oursFullHighlight}

\newcounter{mycomment}

\newcounter{mycommentInline}

\title{\oursTitle}

\author{
    Jongyoon Kim \
    Minseong Hwang \
    Seung-won Hwang\thanks{~~Corresponding Authors}~\\
    Interdisciplinary Program in Artificial Intelligence, Seoul National University\\
    \texttt{\{john.jongyoon.kim, hwmin0823, seungwonh\}@snu.ac.kr} \\
}

\begin{document}
\maketitle

\newabbreviation{uda}{UDA}{Unsupervised Domain Adaptation}
\newabbreviation{ir}{IR}{Information Retrieval}
\newabbreviation{id}{ID}{in-domain}
\newabbreviation{ood}{OOD}{out-of-domain}
\newabbreviation[shortplural={PQs},longplural={pseudo-queries}]{pq}{PQ}{pseudo-query}

\newabbreviation{plm}{PLM}{Pre-trained Language Model}
\newabbreviation{pqg}{PQG}{pseudo-query generator}
\newabbreviation{llm}{LLM}{large language model}
\newabbreviation{mlm}{MLM}{Masked Language Modeling}

\newabbreviation{if}{IF}{Isolation Forest}

\newabbreviation{eig}{EIG}{Expected Information Gain}
\newabbreviation{pu}{PU}{Predictive Uncertainty}
\newabbreviation{au}{AU}{Aleatoric Uncertainty}
\newabbreviation{eu}{EU}{Epistemic Uncertainty}

\begin{comment}
When making decisions under uncertainty, it can be useful
to reason about where that uncertainty comes from (Osband
et al, 2023; Wen et al, 2022). Researchers often aim to do
this by referring to aleatoric (literal meaning: “relating to
chance”) and epistemic (“relating to knowledge”) uncertainty, ideas with a long history in the study of probability
(Hacking, 1975). Aleatoric uncertainty is typically associated with statistical dispersion in data (sometimes thought
of as noise), while epistemic is associated with a model’s internal information state (H ̈ullermeier & Waegeman, 2021).
https://arxiv.org/pdf/2412.20892 (Rethinking Aleatoric and Epistemic Uncertainty ICML2025)
Aleatoric -> Relating to change (Often regarded as a noise)
Epistemic -> Relating to knowledge (associated with the model's internal state)
\end{comment}

\begin{abstract}
Unsupervised domain adaptation generalizes neural retrievers to an unseen domain by generating pseudo queries on target domain documents.
The quality and efficiency of this adaptation critically depend on which documents are selected for pseudo query generation.
The existing document sampling method focuses on diversity but fails to capture model uncertainty.
In contrast, we propose \oursAbbr (\ours) addressing these limitations by (1) filtering documents with high aleatoric uncertainty and (2) prioritizing those with high epistemic uncertainty, maximizing the learning utility of the current model.
We conducted extensive experiments on a large corpus of BEIR with small and large models, showing significant gains of +2.45 and +3.49 nDCG@10 with a smaller training sample size, 4k on average.~\footnote{The implementation is available at: \url{https://github.com/ldilab/UnIte}.
}
\end{abstract}

\glsresetall

\begin{comment}
# 0: 검색모델 좋음 -> 근데 unseen에서 못함 
# 1: UDA가 제안됨 -> 근데 많은 문서에서 PQ 생성 못함.
# 2: 그래서 어떤 문서에서 PQ 만들지 샘플링함. -> 근데 샘플링 방법에 성능이 dependant함.
# 3: 문서 샘플링 방법을 테이블처럼 정리해볼 수 있음. (테이블 없는 게 나으려나..?)
    -> diversity는 도메인에서 다양하게 뽑기
    -> informativeness는 

그냥 active learning으로 framing을 해야하나...

DUQGen의 문제점
높은 다양성을 지향하지만 정작 학습에 어떤 게 가장 유익한지는 모른다.
model-agnostic selection.
1. (문서 뽑는데 타겟 검색 모델의 지식이 아닌 다른 외부 모델의 임베딩으로 뽑음.)
2. 검색 모델을 학습 시킬 문서를 고르는 건데 그 검색 모델에 대한 고려가 없음.

학습시킬 데이터를 고르는데 학습 대상 모델에 대한 고려가 없음.

DUQGen은 Contriever 임베딩 공간에서 coverage+diversity를 최적화하지만, 
    이것이 타겟 모델의 손실 감소나 수렴 가속으로 직결된다는 보장은 없음 (표현 불일치).
DUQGen은 선택→학습이 단발성이라, 학습이 진행되며 변하는 모델의 불확실성/약점 분포를 반영하지 못함(피드백 부재).
큰 클러스터 위주로 샘플이 배정되어 롱테일·경계 사례가 과소대표. 
    랭킹 성능에 임팩트 큰 “희소하지만 어려운” 예제가 빠질 위험.
합성/학습 비용이 유한한 실환경에서, 같은 N이라면 ‘정보성 높은 샘플’에 더 투자해야 데이터 효율성이 올라가는데, DUQGen은 정보성을 직접 다루지 않음.

domain adaptation 의 데이터 구축에서 

DUQGen의 diversity-based sampling은 높은 coverage를 지향하지만
(1) 학습 모델을 고려하지 않아, 그 모델의 성능 개선을 보장하지 않으며, 
(2) 단발성 선택 절차는 변화하는 모델의 불확실성을 반영하지 못한다.
나아가 (3) 높은 diversity를 추구함에 따라 low-density 영역에서 문서를 뽑게 되는 데 이런 문서는 도메인의 대부분의 다른 문서와 다른 단어 분포를 가져 학습에 유용하지 않은 문서를 고르게 된다.

\end{comment}

% \jyin{unified symbol (including styles, mathcal, mathrm etc...)}
% \jyin{abbreviation should be expressed as full words only at first.}
% \jyin{24.15 drop on fig as arrow}

\section{Introduction}
Neural retrievers pre-trained on large datasets, such as MS-MARCO, achieve strong performance in the seen domain while being limited to unseen domains \cite{beir}.
% However, zero-shot performance is often suboptimal for specialized domains. 
\gls{uda} via pseudo query generation has emerged as a promising solution, finetuning the retriever on target-domain documents paired with generated queries \cite{beir,qgen}.
Yet for corpora exceeding 100k documents, the number of generator calls scales with corpus size and is often infeasible under typical budgets~\cite{gospodinov2023doc2query}.
Therefore, sampling \emph{which} documents to query becomes the central bottleneck.

% ========================================================= %

% \begin{wrapfigure}{c}{\textwidth}
\begin{figure}[!t]
{
\centering
    \includegraphics[width=1.0\linewidth]{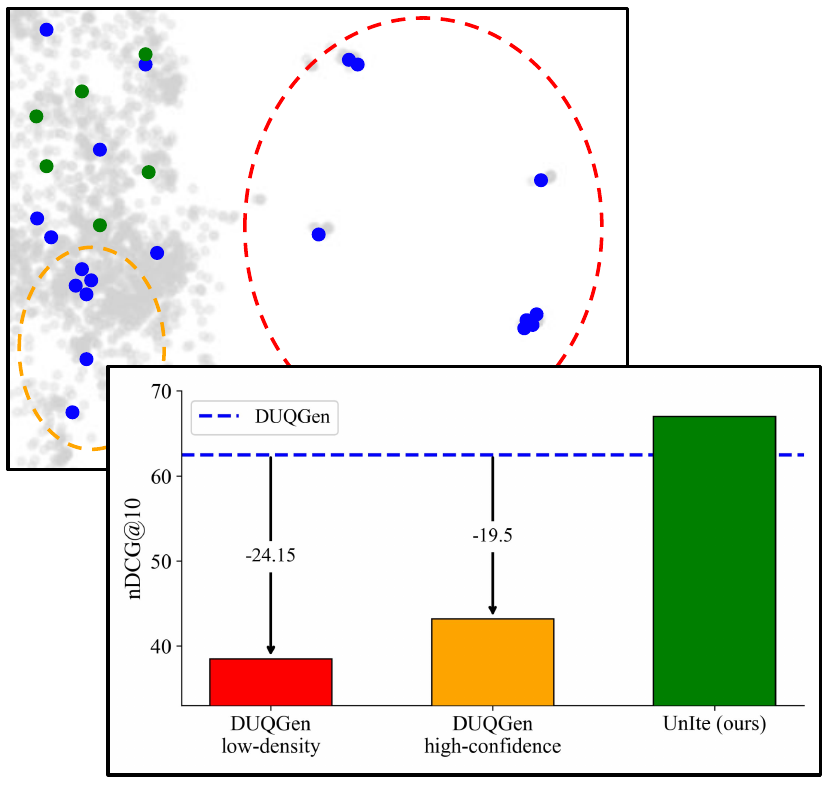}
    \caption{
    The scatter plot shows that DUQGen tends to select samples from low-density (\textcolor{red}{red}) or high-confidence (\textcolor{orange}{orange}) regions in TREC-COVID with Contriever (cropped for visibility). 
    These samples lead to performance degradation, indicating they are suboptimal for domain adaptation.
    }
    \label{fig:introductory}
}
\end{figure}

Unlike prior works that randomly sample documents \cite{gpl,inpars,promptagator}, DUQGen samples with a diversity-based approach \cite{duqgen}, with an external embedding model\footnote{Contriever is used for sampling.}.
While diversity improves coverage, not all regions contribute equally to adaptation. 
We observe two characteristic regions in DUQGen’s sampling as illustrated on \Cref{fig:introductory}: 
(i) \textbf{\textcolor{red}{low-density (red-circled region)}} regions where it contains atypical or outlier documents, and 
(ii) \textbf{\textcolor{orange}{high-confidence (orange-circled)}} regions where the model’s predictive uncertainty is low (already-learned areas), yielding a limited learning signal.

Training exclusively on each region drops performance by  \textcolor{red}{24.15 (red bar)} and \textcolor{orange}{19.32 (orange bar)} nDCG@10 relative to DUQGen (62.56), as shown in \Cref{fig:introductory}, indicating a limited learning signal for adaptation, while sampled heavily (5\% and 13\%).
Meanwhile, according to uncertainty taxonomy \cite{hullermeier2021aleatoric}, \textcolor{red}{low-density regions} correspond to \textcolor{red}{high \gls{au}}, that is inherent in ambiguous or noisy data, while \textcolor{orange}{high-confidence areas} correspond to \textcolor{orange}{low \gls{eu}}.
\gls{eu} reflects the model's knowledge gaps, that is, documents with high \gls{eu} are misaligned with the target domain and thus informative for adaptation.
This suggests prioritizing \emph{low-\gls{au}, high-\gls{eu}} documents while maintaining diversity for coverage.

% proposal still weak?
We propose \oursAbbr \textbf{(\ours)}, which estimates and exploits both forms of uncertainty for document sampling.
(1) \textbf{\gls{au} (data).}
We use a density proxy based on the lexical distance to the $k$-th nearest neighbor. 
If this neighbor is distant, the document is considered to lie in a low-density region.
We filter out noisy documents in low-density regions to prevent selection failures.
(2) \textbf{\gls{eu} (model).}
We measure how well the current retriever aligns a document with the target domain by comparing the document’s embedding with a vocabulary-based domain distribution derived from the model.
A poor document representation of important terms, such as having high frequency in the domain, is treated as high \gls{eu}.

%We then balance uncertainty with diversity during selection, rather than relying on either alone.
Another challenge is that \gls{eu} shifts as the model adapts. 
As a result, good samples 
%Initially informative samples may become trivial, causing adaptation performance to plateau.
may become suboptimal after shifts.
We therefore adopt an \emph{iterative} sample–train loop that recomputes \gls{eu} each round, and we stop when uncertainty plateaus. 
%We exhibit that this correlates with the performance gain in the analysis.

We evaluate on five BEIR datasets with large corpora ($>$100k documents). 
\ours improves over DUQGen by +2.45 on the small model, DPR, and +3.49 on large model, Qwen3-Embedding-4B, on average nDCG@10 while using fewer pseudo-queries. 
Ablations show that removing epistemic sampling drops nDCG@10 by 2.3, and an additional drop of 0.9 is caused by removing the aleatoric filter, underscoring the need for uncertainty-aware selection.

\begin{comment}
* Domain adaptation
* Document sampling
* Our distinction

\end{comment}

\begin{figure*}[!h]
% \begin{wrapfigure}{c}{\textwidth}
{
\centering
    % https://docs.google.com/presentation/d/1oEnBOLN03DNUK1tEU5qJXYZ9_XxNEGqkX8QEe2Zhs7I/edit?slide=id.g3b462a84096_2_525#slide=id.g3b462a84096_2_525
    \includegraphics[width=1.0\textwidth]{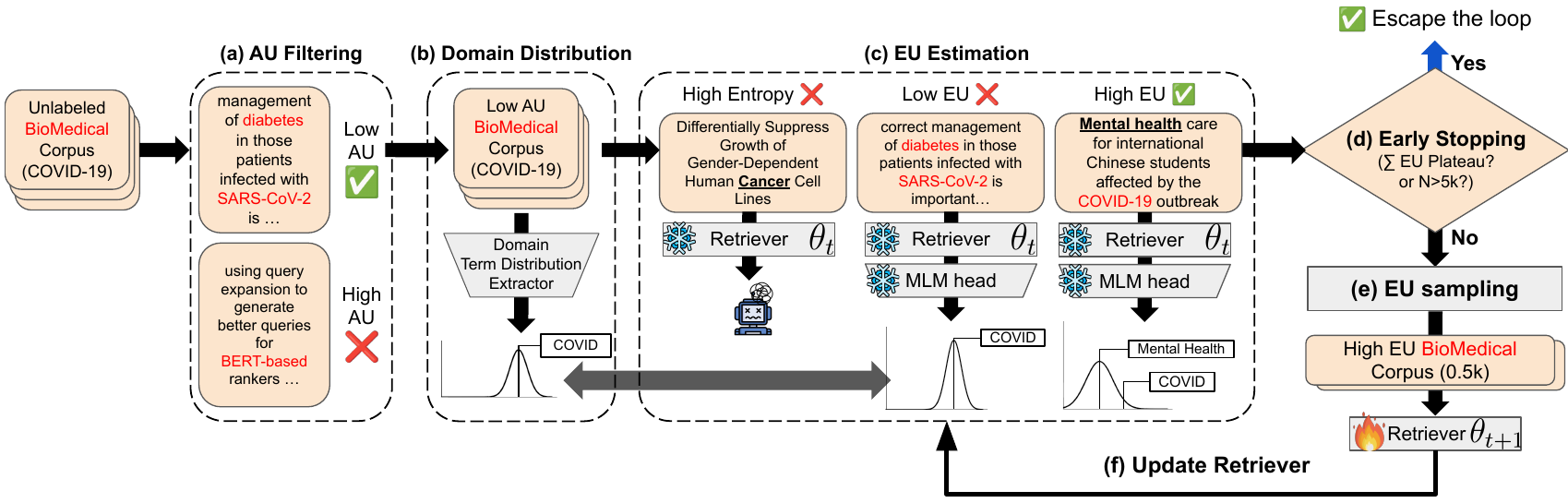}
    \caption{
    \ours pipeline overview. An example from the Biomedical Domain (TREC-COVID) is illustrated. \gls{au} filtering is performed first on the corpus. Then, the sampling-training loop samples documents based on \gls{eu} and iterates until the maximum budget is reached or meets early stopping criteria.
    }
    \label{fig:pipeline}
}
\end{figure*}
% \end{wrapfigure}

\section{Related Work}
This section synthesizes prior \gls{uda} works and studies the progress in document selection for \gls{uda}.
We organize the discussion into three parts: (i) \gls{uda} with pseudo-query generation, (ii) document sampling strategies, and (iii) our distinction.

\paragraph{Unsupervised Domain Adaptation}
\gls{uda} for neural retrieval typically proceeds by generating pseudo-queries from target-domain documents and then fine-tuning a source-trained retriever, such as, one trained with MS-MARCO~\cite {beir,gpl}.
Early approaches used Doc2Query~\cite{gospodinov2023doc2query}, while recent ones employ LLM prompting with domain descriptions~\cite{tart}, contrastive~\cite{inpars} or few-shot examples~\cite{promptagator}, or query expansion~\cite{lee-etal-2025-trag}.
Several approaches also filter low-quality queries using perplexity or consistency checks~\cite{inpars,promptagator}.
These methods improve pseudo-query quality, but generating queries for all documents is infeasible, leading most systems to randomly sample documents under a fixed budget.

% \paragraph{Unsupervised Domain Adaptation}
% \gls{uda} for neural retrieval typically proceeds by generating pseudo-queries from target-domain documents and then fine-tuning a source-trained retriever, that is, one trained with MS-MARCO~\cite {beir,gpl}. 
% Early approaches rely on doc2query~\cite{gospodinov2023doc2query} when generating pseudo-queries and have since evolved into LLM-prompting methods. 
% Such prompting strategies include target-domain descriptions~\cite{tart}, contrastive prompts that show good and bad outputs~\cite{inpars}, few-shot examples per domain~\cite{promptagator}, and LLM-based query term expansion~\cite{lee-etal-2025-trag}. 
% Many works further apply filters based on LLM-derived signals, such as perplexity~\cite{inpars} or consistency checks~\cite{promptagator}, which test whether the generated query is answerable by the given document. 
% These approaches focus on improving the quality of pseudo queries to mitigate the query–document gap.
% However, because retrieval corpora are typically large, generating pseudo-queries for every document is infeasible, as LLM calls scale with corpus size. 
% Consequently, most systems randomly sample documents within a fixed budget, implicitly assuming uniform contribution to target domain adaptation.

\paragraph{Document Sampling}
Because pseudo queries are generated from sampled documents, gains from improving the quality of pseudo queries are fundamentally bounded by the sampling step. 
Consequently, recent \gls{uda} methods seek to improve the quality of the sampled documents themselves. 
Prominent directions in document sampling include 
(i) maximizing target-domain coverage via diversity-based selection~\cite{duqgen} and 
(ii) estimating document quality by training neural models to approximate perplexity or related signals~\cite{qualityscore,generate-distill}. 
% However, these approaches do not account for both the target-domain distribution and the current model knowledge on the target domain, and thus often select uninformative documents for adaptation, 
% thereby wasting the selection budget.
However, these approaches do not account for both the target-domain distribution and the current model knowledge on the target domain.
For example, DUQGen~\cite{duqgen} relies on an external model (Contriever) that is irrelevant to the target retriever's learning state, and performs a single static selection that cannot adapt as the model trains. As a result, it
often selects uninformative documents for adaptation, thereby wasting the selection budget.
% thereby diluting the learning signal under a fixed budget.
% (i) low-density (high-aleatoric) regions that frequently contain off-topic or noisy content, and (ii) low-adaptive (low-epistemic) regions comprising documents the model already knows well, thereby diluting the learning signal under a fixed budget.

\paragraph{Our Distinction}
Unlike prior work DUQGen \cite{duqgen}, which neglects the target-domain distribution and the model’s current knowledge, we explicitly estimate uncertainty at both the data and model levels. 
These estimations drive our sampling policy, which prioritizes low-\gls{au}, high-\gls{eu} documents with diversity to keep coverage, so that the selected set is tailored to the specific domain and retriever. 
This yields more informative updates per query-generation budget than random or purely diversity-based selection.

% \paragraph{Our distinction}  
% In contrast to prior works that (1) sample documents \textit{model-agnostically}, that ignore the retriever’s current embedding space, and (2) generate queries from generic or loosely selected contexts, \ours is the first framework to jointly optimize document selection via an IDF‐weighted \textbf{uncertainty} metric and query generation via an \textbf{information‐density criterion}. 
% By unifying these objectives under a core‐set utility, \ours directly aligns data sampling with the retriever’s semantic gaps, yielding more effective unsupervised adaptation.

% ==================================
\begin{comment}
AU filtering
Train-Sampling loop
 -> eu estimation
 -> early stopping
 -> eu sampling
 --> resampling penalty
 -> 
 
\end{comment}
\section{Methodology}
As shown in \Cref{fig:pipeline}, this section explains how \ours leverages uncertainties. 
The algorithmic details are provided in \Cref{sec:appendix:algo}.

\subsection{\gls{au} Filtering}
\Cref{fig:pipeline}-(a) illustrates a concrete example of this filtering process within the Biomedical (COVID-19) domain. 
The corpus contains some off-topic documents discussing "BERT-based rankers". 
Since this document is lexically distant from the dense region of biomedical terms, \ours identifies its document $d$ as having high \gls{au} and filters it out from the corpus $\mathcal{C}$, thereby preventing negative transfer.

% Following \citet{hacohen2022activelearningbudgetopposite}, we identify outliers in low-density regions as high-\gls{au} samples. 
% To quantify this density independently of the neural model and focus only on data-inherent uncertainty, we adopt a lexical distance metric \citep{hu-etal-2019-domain-adaptation} based on the BM25 score. 

Following \citet{hacohen2022activelearningbudgetopposite}, we identify outliers in low-density regions as high-\gls{au} samples. 
Since \gls{au} is inherently a property of the data itself, independent of any particular model~\citep{hullermeier2021aleatoric}, its estimation must also be model-free. 
Using neural embeddings for this purpose would conflate data uncertainty with model uncertainty. 
A document may appear as an outlier simply because the model has not yet learned to represent it (high \gls{eu}), rather than being inherently noisy (high \gls{au}). 
To ensure clean separation from \gls{eu}, 
we adopt a lexical distance metric \citep{hu-etal-2019-domain-adaptation} based on the BM25 score, which relies solely on corpus statistics.

Formally, the distance to the $k$-th nearest neighbor $n_k$ is defined as: 
\begin{equation}
D_k(d) = \frac{1}{\epsilon + \mathrm{BM25}(d, n_k)}
\end{equation}
where $\epsilon$, set to $1e-6$, prevents zero division. 
We normalize these distances using the modified z-score $z(\cdot)$ to ensure compatibility across domains.
As illustrated in \Cref{fig:pipeline}-(a), off-topic documents, such as the one discussing "BERT-based rankers", exhibit high distances ($z(d) > z_{\mathrm{thr}}$) due to the lack of shared terms with the corpus. 
These high \gls{au} documents are filtered out to yield a refined corpus $\mathcal{C}'$: 
\begin{equation}
\label{eq:zthr}
\mathcal{C}'=\{d \in \mathcal{C} \mid z(d) \le z_{\mathrm{thr}}\}
\end{equation}

% 2. Iterative Sampling-Training Loop
\subsection{Iterative Sampling-Training Loop}
A one-shot sampling strategy relies on a static assessment of document informativeness derived from the initial model $\theta_0$, failing to capture the model's evolving understanding.
Documents that were initially uncertain may become trivial over time, causing static selections to suffer from information redundancy \cite{settles2009active, ash2019deep}.
To address this, we employ an iterative sampling-training loop (\Cref{fig:pipeline}-(c–f)) that continuously re-evaluates \gls{eu} and re-balances domain coverage via dynamic sampling budget allocation based on the updated model state $\theta_t$.

% A one-shot sampling strategy relies on a fixed snapshot of the adaptation dataset constructed with the initial model, failing to capture the model's evolving understanding. 
% As active learning theory demonstrates, document informativeness shifts during adaptation.
% The samples that initially appeared uncertain may become trivial, causing static selections to suffer from information redundancy~\citep{settles2009active, ash2019deep}. 
% Without iterative refinement, the model risks overfitting to initially dominant topics while neglecting evolving decision boundaries. 
% To address this, we employ an iterative sampling-training loop (\Cref{fig:pipeline}-(c–f)) that continuously re-evaluates \gls{eu} and re-balances domain coverage via dynamic sampling budget allocation based on the model's current state.

\paragraph{\gls{eu} Estimation}
The first step in each iteration is to identify documents that bridge the model's current knowledge gap regarding the target domain. 
Prior \gls{eu} estimation, such as Entropy, relies solely on model variance, often failing to detect domain misalignment, as illustrated in \Cref{fig:pipeline}-(c, High Entropy).
In contrast, we measure \gls{eu} by contrasting the model's representation with the target domain statistics.

Prior to the iterative sampling, we pre-compute the target domain statistics, specifically token-level IDF, as illustrated in \Cref{fig:pipeline}-(b).
Then, we project the document embedding $e_d$ to the vocabulary space via the model's MLM head to obtain token probabilities $p(t|e_d; \theta_t)$, see \Cref{fig:pipeline}-(c).
For the document embedding, we follow the model's pooling method, that is, mostly mean pooling for an encoder-based model, and last token pooling for a decoder-based model (details in \autoref{sec:appendix:other-results}.
A document is considered informative (high \gls{eu}) when the model fails to predict high-IDF domain terms, in contrast to low \gls{eu} documents where predictions are confident.
Formally, the \gls{eu} score $U_k(d)$ aggregates the discrepancy between the domain importance (IDF) and the model's prediction for the top-$k$ tokens $T_k(d; \theta_t)$:
\begin{equation}
U_k(d; \theta_t) = \sum_{t \in T_k(d; \theta_t)} [\log \mathrm{IDF}(t) - p(t|e_d; \theta_t)]
\end{equation}
We set $k=1000$, which covers approximately 90\% of the cumulative probability mass (about 3\% of BERT's 32k vocabulary).

Crucially, as the model adapts to the target domain, $p(t|e_d; \theta_t)$ evolves, dynamically altering the set of high \gls{eu} documents in each round.

% We limit the comparison to top-$k$ tokens for tractable calculation.
% Because tokens with high IDF that are predicted with low probability contribute most to $U_k(d; \theta_t)$, this score effectively flags documents exhibiting a knowledge gap in the model.

% 2-2. Early Stopping
\paragraph{Early Stopping Criteria}
Once the uncertainty scores are calculated, determining when to stop is crucial to prevent overfitting and save costs. 
As depicted in \Cref{fig:pipeline}-(d), we hypothesize that the domain-averaged \gls{eu} reflects the model's knowledge saturation. 
We monitor this score using an Exponential Moving Average (EMA) to smooth out fluctuations. 
The loop terminates when the smoothed \gls{eu} reaches a plateau (local minimum), or the maximum budget (5k) is exhausted. 
As validated in our analysis, this unsupervised criterion effectively signals the point of peak retrieval performance.

% 2-3. EU Sampling
\paragraph{\gls{eu} Sampling}
If the stopping criteria are not met, we proceed to sample documents for adaptation training.
While uncertainty sampling targets knowledge gaps, it implies a risk of sampling redundant documents within high-uncertainty regions. 
To ensure coverage across the domain, we adopt the diversity-driven clustering framework of DUQGen~\cite{duqgen}. 
However, unlike the baseline, which relies solely on diversity, we balance it with our estimated uncertainty.
Specifically, within each semantic cluster $\mathcal{C}_i$, we prioritize documents using a Maximal Marginal Relevance (MMR) approach~\cite{mmr}. 
We select the top-$n_i$ documents that maximize the following joint score within each cluster for the current retriever model state $\theta_t$:
\begin{equation}
\label{eq:score-balance}
    \text{score}(d; \theta_t)
= \lambda\,\widehat{\mathrm{U}_k(d; \theta_t)} + (1-\lambda)\,\widehat{\Psi(d; \theta_t)}
\end{equation}
where $n_i$ indicates the sampling budget for the $i$-th cluster, $\widehat{\cdot}$ denotes z-score normalization, and $\Psi$ is the diversity score from DUQGen. 
This strategy ensures that the final training set $S = \bigcup S_i$ is composed of documents that are both informative (high \gls{eu}) and representative of diverse topics (high diversity).

\begin{comment}
Another challenge is that \gls{eu} shifts as the model adapts. 
As a result, good samples may become suboptimal after shifts.
We therefore adopt an \emph{iterative} sample–train loop that recomputes \gls{eu} each round, and we stop when uncertainty plateaus. 

    문서에 대한 eu 가 학습함에 따라 변화하게됨.
    학습에 활용한 문서의 EU는 학습 후 EU가 낮아지게 됨.
    즉, 그 데이터는 아무리 좋은 문서라 할지라도, 더이상 adaptation으로써의 효용이 낮아짐.
    그렇기 때문에, iterative하게 sampling-training을 해야함.
\end{comment}

\paragraph{Addressing \gls{eu} Shift via Iteration}
A critical challenge in domain adaptation is the dynamic nature of \gls{eu}, which evolves as the model trains.
According to active learning theory \cite{settles2009active}, the \gls{eu} of documents shifts throughout the training process, gradually decreasing in regions where the model has already been exposed to training samples.
Consequently, clusters that initially exhibited high \gls{eu} progressively lose their informativeness for further adaptation to the target domain. 
This necessitates an iterative sampling that actively avoids redundant selection from regions exhibiting decreased \gls{eu}. 
Such an approach ensures that subsequent sampling iterations continuously target the model's evolving knowledge gaps in the target domain rather than repeatedly sampling from clusters where the model has already trained.

\paragraph{Resampling Penalty for Iterative Sampling}
Prior approaches like DUQGen, however, allocate the sampling budget $n_i$ in proportion to the static cluster size $|\mathcal{C}_i|$, without accounting for the evolving nature of \gls{eu}. 
This static allocation strategy neglects \gls{eu} shifts across training, resulting in the model overfitting to the dominant clusters while neglecting minority clusters where knowledge gaps persist.

To mitigate this limitation, we introduce a resampling penalty that dynamically redistributes sampling attention toward underrepresented clusters exhibiting high \gls{eu} (\Cref{fig:pipeline}-(e)). 
Specifically, the sampling weight $w_i$ for the $i$-th cluster is adjusted inversely to its accumulated sample count $\mathcal{P}_i$ over previous iterations:
\begin{equation}
\label{eq:penalty}
w_i = \frac{|\mathcal{C}_i|}{\mathcal{P}_i + \epsilon}, \quad \text{where} \quad n_i = n \cdot \frac{w_i}{\sum_j w_j}.
\end{equation}
By penalizing redundant sampling from previously explored dominant clusters, this mechanism progressively shifts sampling weight toward underrepresented minority regions.

\section{Experimental Setup}
\begin{table*}[t]
  \centering
  \small
  \scalebox{0.95}{
  \renewcommand{\arraystretch}{1.05}
  \begin{tabular}{ 
      l 
      l 
      || c c c c c
      || c 
    }
    \toprule
    \multicolumn{1}{c}{
        \multirow{2}{*}[-0.8ex]{\textbf{Retriever}}
    }
      & \multicolumn{1}{c||}{
            \multirow{2}{*}[-0.8ex]{
            \makecell[c]{\textbf{Adaptation}\\\textbf{Method}}}
      } 
      & \multicolumn{5}{c||}{\textbf{Large Corpus}} 
      & \multicolumn{1}{c}{
            \multirow{2}{*}[-0.8ex]{\textbf{Total AVG}}
      } \\
    \cmidrule(lr){3-7}
      & 
      % & \textbf{FQ} 
      & \textbf{TC} & \textbf{RB} & \textbf{QR} 
      & \textbf{TN} 
      % & \textbf{CQA} & \textbf{NQ} 
      & \textbf{HQ} \\ 
    % \midrule
    % \multicolumn{12}{l}{\textbf{First-stage Retriever}} \\
    \midrule
    BM25
      & —      
      % & 23.61 
      & 65.59 & 40.70 & 78.9 
      & 39.8 & 60.3
      % & 29.9
      % & 32.9 
      & 44.49 \\
    \midrule
    \multirow{5}{*}{DPR}  
      & — $\dag$
      % & 11.2 
      & 33.2 & 25.2 & 24.8
      & 16.1 & 39.1
      % & 15.3 & 47.4 
      & 27.68 \\
      & Random  
      & 60.81 & 31.11 & 74.01
      & \underline{28.66} & 38.02
      & 46.52 \\

      % \revision{
          
      % }
      % & GPL 
      % & \underline{63.36} & \underline{31.90} & QR
      % & TN & HQ
      % & AVG \\

      & Quality  
      & 61.11 & 28.25 & 75.01
      & 23.42 & 39.35
      & 45.43 \\
      & DUQGen  
      & 62.75 & 31.48 & \underline{75.17} 
      & 24.05 & \underline{39.62}
      & \underline{46.61} \\
      & \ours
      & \textbf{66.79$\flat$} \textcolor{brown}{$\uparrow$4.04} & \textbf{33.23*} \textcolor{brown}{$\uparrow$1.75} & \textbf{75.32} \textcolor{brown}{$\uparrow$0.15}
      & \textbf{29.13*} \textcolor{brown}{$\uparrow$5.08} & \textbf{40.82*} \textcolor{brown}{$\uparrow$1.20}
      & \textbf{49.06$\flat$} \textcolor{brown}{$\uparrow$2.45} \\
    % \midrule
    % \multirow{3}{*}{ColBERT}  
    %   & — $\dag$
    %   & 70.6 & 39.2 & 85.3
    %   & 39 & 59
    %   & 58.62 \\
    %   & DUQGen  
    %   & \textbf{74.18} & 44.95 & \textbf{85.57}
    %   & 36.93 & -
    %   & - \\
    %   & OASiS   
    %   & 73.43 & \textbf{46.37} & 85.09
    %   & \textbf{37.48} & \textbf{-}
    %   & \textbf{-} \\
    \midrule
    
    \multirow{5}{*}{coCondenser}  
      & — 
      & 67.48 & \underline{32.51} & 86.36
      & 28.9 & 54.44
      & 53.94 \\
      & Random  
      & 66.35 & 32.26 & \underline{87.12}
      & 28.14 & 56.00
      & 53.97 \\
      & Quality  
      & 63.67 & 31.36 & 86.61
      & \textbf{31.95} & 56.17
      & 53.95 \\
      & DUQGen  
      & \underline{70.35} & 32.42 & \underline{87.12}
      & 28.02 & \textbf{56.77}
      & \underline{54.94} \\
      & \ours
      & \textbf{71.02} \textcolor{brown}{$\uparrow$0.67} & \textbf{33.95*} \textcolor{brown}{$\uparrow$1.53} & \textbf{87.17} \textcolor{brown}{$\uparrow$0.05} 
      & \underline{31.16*} \textcolor{brown}{$\uparrow$3.14} & \underline{55.14} \textcolor{brown}{$\downarrow$1.63}
      & \textbf{55.69$\flat$} \textcolor{brown}{$\uparrow$0.75} \\
    \midrule

    \multirow{6}{*}{COCO-DR}  
      & — 
      & \underline{79.34} & 44.64 & 86.73
      & \textbf{38.61} & 60.43
      & 61.95 \\
      & Random  
      & 79.18 & 44.8 & 87.05
      & \underline{38.34} & 60.53
      & 61.98 \\
      & GPL  
      & 78.38 & 43.72 & 86.95
      & 37.3 & 59.82
      & 61.23 \\
      & Quality  
      & 79.12 & \underline{45.09} & 86.88
      & 37.96 & \underline{60.6}
      & 61.93 \\
      & DUQGen  
      & 79.16 & 45.05 & \underline{87.15}
      & 37.84 & \textbf{60.84}
      & \underline{62.01} \\
      & \ours
      & \textbf{80.02*} \textcolor{brown}{$\uparrow$0.86} & \textbf{45.29} \textcolor{brown}{$\uparrow$0.24} & \textbf{87.16} \textcolor{brown}{$\uparrow$0.01} 
      & 38.31$\flat$ \textcolor{brown}{$\uparrow$0.47} & 60.56 \textcolor{brown}{$\downarrow$0.28}
      & \textbf{62.27$\flat$} \textcolor{brown}{$\uparrow$0.26} \\
      \midrule

    \multirow{4}{*}{Qwen3 (4B)}  
      & — 
      & \underline{88.91} & \underline{62.27} & \underline{88.30}
      & 21.17 
      
      & -
      % & HQ
      
      & 65.16 \\
      & GPL  
      & 88.81 & 59.84 & 88.28
      & 21.89 & -
      & 64.71 \\
      & DUQGen  
      & 88.60 & 60.90 & 83.90
      & \underline{43.82} 
      
      & -
      % & HQ
      
      & \underline{69.31} \\
      & \ours
      & \textbf{91.60*} \textcolor{brown}{$\uparrow$3.00} & 
      \textbf{62.31*} \textcolor{brown}{$\uparrow$1.41} & 
      \textbf{88.62*} \textcolor{brown}{$\uparrow$4.72} & 
      \textbf{48.68*} \textcolor{brown}{$\uparrow$4.86} 
      & -
      % & \textbf{HQ} \textcolor{brown}{$\uparrow$HQdelta} 
      & \textbf{72.80*} \textcolor{brown}{$\uparrow$3.49} 
      \\
      
    % \revision{
        
    % }
    % \midrule
    % \midrule\midrule
    % \multicolumn{12}{l}{\textbf{Reranking BM25 Top-100}} \\
    % \midrule
    % \multirow{4}{*}{monoT5} 
    %   & — $\dag$      
    %   & 73.04 & 35.71 & 18.83 & 17.10 
    %   & 36.17 & 39.10 & 81.39 & 51.81 \\
    %   & InPars† 
    %   &   —   &   80.3   &   51   &   —   
    %   &   —   & — & 31.3 & — \\
    %   & DUQGen  
    %   & 39.02 & 84.42 & 54 & \textbf{88.04} 
    %   & 46 & 40.16 & \textbf{54.74} & 58.05 \\
    %   & RaDSenSe   
    %   & \textbf{39.41} & \textbf{84.94} & \textbf{54.2} & 87.97 & \textbf{46.47} & \textbf{40.32} & 54.35 & \textbf{58.24} \\

    \bottomrule
  \end{tabular}
}
  \caption{
  Retrieval performance (nDCG@10) on BEIR across retrievers and adaptation methods. 
  \textbf{Bold} entries mark the highest performances per dataset for each retriever, while the \underline{underlined} entries indicate the second-highest. 
  "AVG" columns report the overall average. 
  $\dag$ indicates values taken from the original paper. 
  \textcolor{brown}{$\uparrow$} is the difference between \ours and DUQGen, and statistically significant improvement over DUQGen is denoted as $*$ ($p<0.05$) and $\flat$ ($p<0.1$).
  }
  \label{tab:main-results}
\end{table*}

\subsection{Implementation Details}
\paragraph{\gls{au} Filtering}
We utilize PySerini's prebuilt BM25 index to calculate the lexical $k$-NN distance for filtering low-density documents. 
We set the neighbor count $k=3$ and the z-score threshold $Z_{thr}=1.5$ to flag and exclude outliers in a domain-adaptive manner.

\paragraph{Iterative Sampling-Training Loop}
For \gls{eu} estimation, we compute probabilities using the top-1,000 tokens and set the balance weight $\lambda=0.5$ to equally weigh uncertainty and diversity. 
In the iterative loop, we sample 500 documents per iteration up to a maximum of 10 iterations (total 5k budget), matching the baseline's scale. 
The early stopping mechanism employs an EMA with a smoothing factor $\alpha=0.4$.

\paragraph{Pseudo-Query Generation}
We employ Llama3-8B-Instruct \cite{llama3} with the template shown in \Cref{sec:appendix:prompt}. 
We generate one query per document using temperature 0.8 and top-p 0.9.

\subsection{Datasets and Evaluation Metrics}
We evaluate our method on five large-scale BEIR datasets ($>$100k documents) where effective selection is challenging: TREC-COVID (TC), Robust04 (RB), TREC-NEWS (TN), Quora (QR), and HotpotQA (HQ). 
Dataset statistics are detailed in \Cref{tab:dataset}. 

To evaluate the retrieval performance of the adapted model, we use normalized Discounted Cumulative Gain (nDCG@10), which measures the number and order of relevant documents ranked in the top-10 retrieved documents.

\subsection{Baselines}
We compared \ours with four document selection strategies under a fixed budget of 5k documents:
\begin{itemize}
    \item Random: the documents are randomly sampled.
    \item GPL \cite{gpl}: the documents are randomly sampled, and the relevance annotations between pseudo query and document are labeled with an expensive cross-encoder. For fair comparison, we adapt it to generate 5 k training samples.
    \item Quality \cite{qualityscore}: the documents are sampled by a neural quality estimator model.
    \item DUQGen \cite{duqgen}: the diversity-based sampling via clustering with external embedding model, Contriever \cite{izacard2022contrastive}.
\end{itemize}

% to place on the page of section 5.
\begin{figure*}[!t]
{
\centering
    \includegraphics[width=0.7\textwidth]{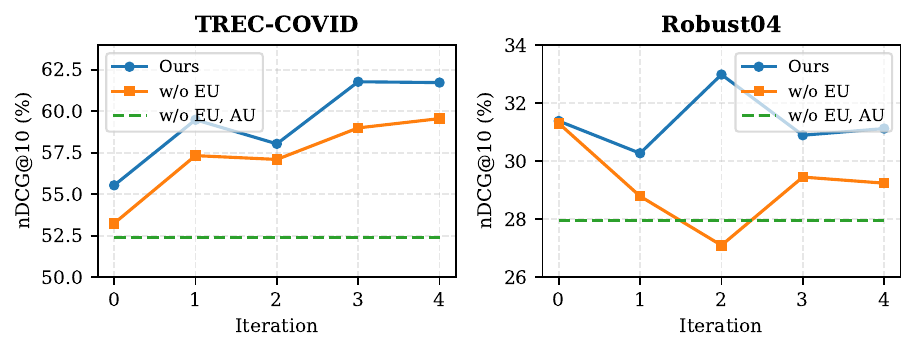}
    \caption{
        Ablation results in an iterative sampling-training loop with and without \gls{au} and \gls{eu}.
        Without both stages, we only measure the first-iteration performance and plot it as a horizontal line.
        We report DPR performance in nDCG@10 on TREC-COVID and Robust04.
    }
    \label{fig:ablation-eu-au}
}
\end{figure*}

\subsection{Retrieval Models}
We assess performance using four single-vector retrievers initially trained on MS-MARCO \cite{msmarco}: DPR \cite{dpr}, coCondenser \cite{gao2021unsupervisedcorpusawarelanguage}, COCO-DR \cite{cocodr}, and Qwen3-Embedding-4B \cite{zhang2025qwen3} \footnote{Additionally, we experiment with ColBERT \cite{colbert} and MonoT5 \cite{monot5} (results in \Cref{sec:appendix:other-results}).}
All models are fine-tuned on the selected documents using their standard objectives. 
Note that all experiments were conducted on a single NVIDIA 3090 GPU, and the detailed configurations are described in \Cref{sec:appendix:fine-tuning}.

% \subsection{Computational Efficiency}
% \revision{We compared the computational cost of \ours against GPL. \ours requires significantly less training time because it selects a small, informative subset (5k documents) compared to GPL's generation on the entire corpus or larger subsets. Specifically, \ours completes the adaptation pipeline in 1.5 hours on a single 3090 GPU, whereas GPL requires over 12 hours for the same scale, highlighting our method's efficiency.}

\section{Results}

\subsection{Overall adaptation performance}
Our results in \Cref{tab:main-results} show that \ours consistently improves nDCG@10 across retrievers and document selection methods, while Random yields occasional gains and Quality sometimes degrades performance.
On average, \ours improves nDCG@10 by +2.45, +0.75, and +0.26 points over DUQGen with DPR, coCondenser, and COCO-DR, respectively.
For DPR, \ours outperforms GPL by a large margin, highlighting the importance of uncertainty-based sampling with limited training data.
With Qwen3-embedding-4B, gains increase to +3.49 points, demonstrating effective scaling with model capacity (HQ excluded as Qwen3-embedding was trained on it \cite{zhang2025qwen3}).
These results confirm that uncertainty combined with diversity yields the most robust adaptation gains.
% In addition, we observe an interesting relationship between the corpus and the performance.
% The corpora with relatively rich document-query distributions, such as TC and QR, benefit noticeably from adaptation, especially with \ours.
% By contrast, on small or highly skewed corpora, like RB and TN, the adaptation results vary widely across methods. That is, the selection method can help one corpus, can hurt the others.
% These findings suggest two points: (1) the choice of adaptation data is highly sensitive to both the retriever and domain characteristics, and (2) \ours appears relatively robust to query variability and domain shift.

\subsection{Ablation Study}
There are two major components in \ours that correspond to uncertainty taxonomy: (i) \emph{\gls{au} Filtering} and (ii) \emph{\gls{eu} Sampling}. 
To empirically prove that both components complementarily contribute to the performance gain, we performed an ablation study in \Cref{fig:ablation-eu-au}. 
We additionally examine the effect of the resampling penalty, which governs budget allocation within the iterative sampling loop, in \Cref{tab:resampling}.
% There are two major components in \ours that correspond to uncertainty taxonomy: (i) \emph{\gls{au} Filtering} and (ii) \emph{\gls{eu} Sampling}.
% To empirically prove that both components complementarily contribute to the performance gain, we performed an ablation study in \Cref{fig:ablation-eu-au}.

\paragraph{Impact of \textit{\gls{eu} Sampling}} 
First, we compared the performance with and without \gls{eu} sampling with DPR on TREC-COVID and Robust04. 
\Cref{fig:ablation-eu-au} shows that \ours (\textcolor{blue}{blue line}) consistently outperforms the baseline without \gls{eu} sampling (\textcolor{orange}{orange line}, "w/o EU") across all iterations on both datasets. 
Notably, on Robust04, the performance without \gls{eu} sampling drops significantly—by approximately 4 nDCG@10 points—compared to the zero-shot. 
This performance gap indicates that document sampling without considering the model's understanding leads to redundant training samples.

\paragraph{Impact of \textit{\gls{au} Filtering}} 
To further assess the contribution of \gls{au} filtering, we conducted an additional experiment by removing this component. 
Without both modules (\textcolor{Green}{green dashed line}, "w/o EU, AU"), we sampled all 5k samples at once, as the method cannot iterate without these components. 
This configuration shows a substantial performance gap of approximately 5 and 9 nDCG@10 points compared to the peak performance of \ours on TREC-COVID and Robust04, respectively. 
This demonstrates that documents sampled from low-density regions critically contribute to false positives when \gls{au} filtering is absent.

\paragraph{Impact of Resampling Penalty}
\begin{table}[t]
\centering
\small
\begin{tabular}{lccc}
\toprule
Setting & TC & QR & TN \\
\midrule
w/ Resampling Penalty & \textbf{61.73} & \textbf{74.95} & \textbf{30.39} \\
w/o Resampling Penalty & 54.39 & 73.42 & 21.83 \\
\midrule
$\Delta$ & +7.34 & +1.53 & +8.56 \\
\bottomrule
\end{tabular}
\caption{Impact of resampling penalty \autoref{eq:penalty} on nDCG@10 with DPR training with 2.5k samples.}
\label{tab:resampling}
\end{table}
To assess the contribution of the resampling penalty \Cref{eq:penalty}, we conducted an ablation study using DPR, and for fair comparison, we fixed the sample size to 2.5k.
As shown in \Cref{tab:resampling}, removing the penalty consistently degrades performance across datasets by about 1.5 to 8.5 nDCG@10, confirming that dynamic budget redistribution prevents over-sampling from dominant clusters.

Hence, removing either of the \gls{au} and \gls{eu} modules harms adaptation, and further performance improvement is largely unattainable. 
Moreover, the resampling penalty complementarily works to ensure balanced coverage across clusters, preventing the iterative loop from  oversampling dominant regions.

\begin{table*}[t]
  \centering
  \small
  \scalebox{1.0}{
  \renewcommand{\arraystretch}{1.1}
  \begin{tabular}{ 
      l 
      l 
      || c c c c c
      || c 
    }
    \toprule
    \multicolumn{1}{c}{
        \multirow{2}{*}[-0.8ex]{\textbf{Retriever}}
    }
      & \multicolumn{1}{c||}{
            \multirow{2}{*}[-0.8ex]{
            \shortstack[c]{\textbf{Adaptation}\\\textbf{Method}}}
      } 
      & \multicolumn{5}{c||}{\textbf{Large Corpus}} 
      & \multicolumn{1}{c}{
            \multirow{2}{*}[-0.8ex]{\textbf{Total AVG}}
      } \\
    \cmidrule(lr){3-7}
      & 
      % & \textbf{FQ} 
      & \textbf{TC} & \textbf{RB} & \textbf{QR} 
      & \textbf{TN} 
      % & \textbf{CQA} & \textbf{NQ} 
      & \textbf{HQ} \\ 
    % \midrule
    % \multicolumn{12}{l}{\textbf{First-stage Retriever}} \\
    % \midrule
    % BM25
    %   & —      
    %   % & 23.61 
    %   & 65.59 & 40.70 & 78.9 
    %   & 39.8 & 60.3
    %   % & 29.9
    %   % & 32.9 
    %   & 44.49 \\
    \midrule
    \multirow{2}{*}{DPR}  
      % & — $\dag$
      % % & 11.2 
      % & 33.2 & 25.2 & 24.8
      % & 39.8 & 39.1
      % % & 15.3 & 47.4 
      % & 28.83 \\
      & DUQGen 
      & 5.91 & 1.26 & 10.07
      & 1.59 & 0.10
      & 3.56 (5k) \\
      & \ours
      & \textbf{9.6} (3.5k) & \textbf{1.61} (5k) & \textbf{10.1} (5k)
      & \textbf{3.26} (4k) & \textbf{0.34} (5k)
      & \textbf{4.50} (4.5k)  \\
    \midrule
    \multirow{2}{*}{coCondenser}  
      % & — $\dag$
      % & 67.48 & \underline{32.51} & 86.36
      % & \underline{28.9} & 54.44
      % & 53.94 \\
      & DUQGen  
      & 0.57 & -0.02 & 0.15
      & -0.18 & \textbf{0.47}
      & 0.20 (5k) \\
      & \ours
      & \textbf{1.18} (3k) & \textbf{0.29} (5k) & \textbf{0.2} (4k)
      & \textbf{0.5} (4.5k) & \underline{0.14} (5k)
      & \textbf{0.41} (4.3k) \\
    \midrule
    \multirow{2}{*}{COCO-DR} 
      % & — $\dag$
      % & \underline{79.34} & 44.64 & 86.73
      % & \textbf{38.61} & 60.43
      % & 61.95 \\
      & DUQGen  
      & -0.04 & 0.08 & 0.08
      & -0.15 & \textbf{0.08}
      & 0.01 (5k) \\
      & \ours
      & \textbf{0.14} (5k) & \textbf{0.13} (5k) & \textbf{0.09} (5k) & \textbf{-0.06} (5k) & 0.03 (5k) & \textbf{0.06} (5k) \\
    \midrule
    \multirow{2}{*}{Qwen3-Embedding-4B} 
      % & — $\dag$
      % & \underline{79.34} & 44.64 & 86.73
      % & \textbf{38.61} & 60.43
      % & 61.95 \\
      & DUQGen  
      & -0.06 & -0.27 & -0.88
      & 4.53 & -
      & 0.83 (5k) \\
      & \ours
      & \textbf{0.54} (5k) & \textbf{0.01} (5k) & \textbf{0.06} (5k) & \textbf{9.17} (3k) & - & \textbf{2.45} (4.5k) \\
    \bottomrule
  \end{tabular}
}
  \caption{
  Retrieval performance gain over zero-shot for unit sampling size ($\Delta$ nDCG@10 / 1k).
  Total sampling size is shown in brackets, except for DUQGen, which always samples 5k.
  \textbf{Bold} values indicate better performance per dataset for each retriever.
  }
  \label{tab:per-query-ndcg-gain}
\end{table*}
 % early stopping

\subsection{\gls{eu} estimation methods}
\label{sec:appendix:eu-estimation}
\begin{table}[h]
\centering
\scalebox{0.9}{
\def\arraystretch{1.05}
% ALL DATASET
% \begin{tabular}{@{}l|ccccccc@{}}
% \toprule
%                                         & SF    & NF    & ARG   & SD    & FQ    & TC    & RB    \\ \midrule
% \textbackslash{}ours                    & 38.54 & 24.19 & 41.87 & 9.63  & 15.79 & 60.73 & 30.04 \\
% - informative span guided PQgen         & 44.36 & 22.99 & 40.7  & 9.27  & 14.83 & 63.68 & 32.90 \\
% - domain uncertainty document selection & 40.73 & 22.93 & 38.51 & 10.25 & 14.97 & 51.63 & 28.37 \\ \bottomrule
% \end{tabular}

% PARTIAL
\begin{tabular}{@{}l|ccc|c@{}}
\toprule
                      & TC   & RB    & TN  & AVG  \\ \midrule
\ours                & \textbf{55.54} & \textbf{31.38} & \textbf{23.33} & \textbf{36.75} \\
MC-Dropout            & 52.79  & \underline{28.27} & 21.6 & \underline{34.22} \\
Entropy              & \underline{54.1} & 25.83 & \underline{22.7} & 34.21 \\ \bottomrule
\end{tabular}
}
\caption{
% Comparison between other uncertainty measures and \ours. 
% DPR performance on the first iteration is reported on three datasets (TC, RB, TN) and their average (AVG). 
% The first row shows \ours uncertainty. The best result for each column is highlighted in \textbf{bold} and the second-best result is underlined.
Comparison of uncertainty measures with \ours. 
DPR performance (nDCG@10) on the first iteration is reported on TC, RB, TN, and their average (AVG). 
The best result for each column is highlighted in \textbf{bold} and the second-best result is \underline{underlined}.
}
\label{tab:compare-eu-algos}
\end{table}
In prior work, \gls{eu} has predominantly been discussed solely in terms of the model itself.
However, in \gls{uda}, it is important to measure the model understanding of \emph{target domain} to properly estimate the \gls{eu}, by introducing distribution statistics.

To assess the effectiveness of incorporating the target domain distribution into \gls{eu} estimation for \gls{uda}, we evaluated adaptation performance under a fixed sampling strategy while varying the estimation methods. 
Specifically, we conducted experiments on TC, RB, and TN with DPR, using a budget of 500 documents.
\Cref{tab:compare-eu-algos} shows that our \gls{eu} estimation method using target domain distribution consistently outperforms other estimation methods across all three domains, yielding an average improvement of 2.53 nDCG@10, over MC-Dropout. 
Since entropy and MC-dropout \gls{eu} estimation overlook target-distribution signals, their estimations are consequently less domain-aware, which results in degraded retrieval performance after adaptation.
Overall, our findings explicitly demonstrate that effective document selection for domain adaptation requires incorporating the target domain distribution into \gls{eu} estimation, which enables \gls{eu} to focus on more valuable samples.

\subsection{Early Stopping Criteria}
\begin{figure}[h]
{
\centering
    \includegraphics[width=0.95\linewidth]{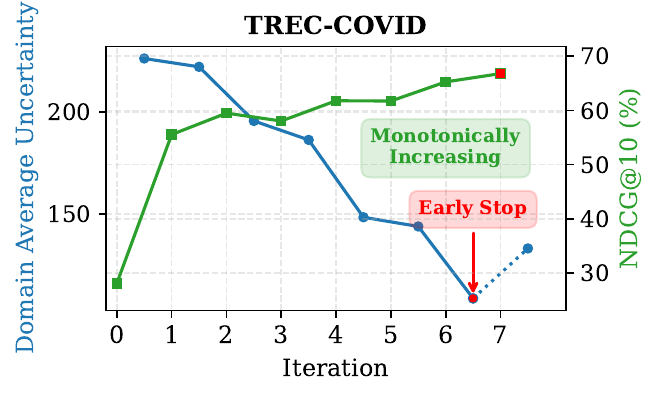}
    \caption{
        Relationship between model performance and average uncertainty across the iterative sampling-training loop.
        Average \gls{eu} across the target domain (\textcolor{blue}{left axis}) and nDCG@10 (\textcolor{Green}{right axis}) is reported on TREC-COVID.
        % The elbow—where uncertainty bottoms out and begins to rise—coincides with the peak nDCG@10, providing a practical early-stopping point.
    }
    \label{fig:early-stopping}
}
\end{figure}

\label{sec:res:early-stopping}
While increasing the number of training documents may appear beneficial, prior work has demonstrated that excessive samples can degrade performance through overfitting~\cite{duqgen}.
As illustrated in \Cref{fig:early-stopping}, the average \gls{eu} initially decreases as the model adapts to the target domain, then subsequently increases, indicating sample redundancy. 
Notably, the local minimum in average \gls{eu} coincides with peak retrieval performance, validating EU as a reliable early stopping criterion.
This characteristic enables \ours to achieve both effectiveness and sample efficiency.
As demonstrated in \Cref{tab:per-query-ndcg-gain}, \ours achieves a 0.94 nDCG@10 improvement per 1k samples with DPR compared to DUQGen, and typically converges at 3--5k samples, in contrast to DUQGen, which utilizes 5k samples.

\subsection{Computational Cost}
The sample efficiency achieved through early stopping directly reduces computational costs through smaller training datasets and fewer adaptation iterations.

\paragraph{Training time}
Training DPR with the baseline 5k dataset requires approximately 10 minutes on a single NVIDIA 3090 GPU, whereas ours utilizes early stopping, which typically produces 3-5k samples, completing adaptation in around 8 minutes.

\paragraph{Dataset construction time}
For the domain adaptation method, most of the additional computation is due to constructing the adaptation dataset.
For our method, \gls{au} filtering is applied once on the full corpus $\mathcal{C}$, requiring 120 seconds, and \gls{eu} estimation operates on the filtered corpus $\mathcal{C'}$ at each iteration, requiring 150 seconds per iteration. 
Since \gls{eu} scales linearly with corpus size, the overhead remains practical even for larger corpora.
The total overhead is $120 + 150 \times N$ seconds, where $N$ is the number of iterations until convergence.
As our method early stops at 4k samples on average, the net time savings are $880$ seconds\footnote{$2.2 \times 1000 - 120 - 150 \times 8$}.
Even with 5k samples, the overhead remains modest at approximately 20 minutes, demonstrating reasonable computational cost for superior performance.

\subsection{Hyperparameter Tuning}
\label{sec:appendix:hyperparameters}
\begin{table}[t]
    \centering
    \footnotesize
    \setlength{\tabcolsep}{3pt}
    \begin{subtable}[t]{0.48\linewidth}
        \centering
        \caption{Impact of $Z_{thr}$}
        \begin{tabular}{l c c c}
        \toprule
        Method & $Z_{thr}$ & nDCG & Ratio \\
        \midrule
        Zeroshot & - & 28.58 & - \\
        \midrule
        \multirow{5}{*}{\shortstack{\ours \\ w/o \gls{eu}}} & 1.0 & 28.57 & 10\% \\
         & \textbf{1.5} & \textbf{29.27} & 5\% \\
         & 2.0 & 28.82 & 3.8\% \\
         & 2.5 & 28.82 & 1.2\% \\
         & 3.0 & 27.47 & 0.4\% \\
        \bottomrule
        \end{tabular}
    \end{subtable}
    \hfill
    \begin{subtable}[t]{0.48\linewidth}
        \centering
        \caption{Impact of $\lambda$}
        \begin{tabular}{l c c}
        \toprule
        Method & $\lambda$ & nDCG \\
        \midrule
        DUQGen & - & 15.37 \\
        \midrule
        \multirow{5}{*}{\ours} & 0.1 & 15.37 \\
         & 0.3 & 14.16 \\
         & \textbf{0.5} & \textbf{16.32} \\
         & 0.7 & 15.32 \\
         & 0.9 & 15.33 \\
        \bottomrule
        \end{tabular}
    \end{subtable}
    \caption{Hyperparameter tuning. Left: Impact of $Z_{thr}$ (CoCondenser). Right: Impact of $\lambda$ (DPR).}
    \label{tab:hyperparam_sensitivity}
\end{table}

We conducted hyperparameter tuning on FiQA, the largest dataset under 100k documents in BEIR, and applied the selected values across all other datasets.
Note that tuning directly on target domains would constitute information leakage in our out-of-domain evaluation setting. 
Therefore, we selected FiQA to serve as a held-out development set. 
As shown in Table 1, the fixed hyperparameters yield consistent improvements across all five evaluation datasets, supporting their cross-domain robustness.
% As our experimental setting targets large corpora ($>$100k documents) in BEIR without dedicated development sets, FiQA serves as a representative validation benchmark.

\paragraph{Filtering Threshold ($\mathcal{Z}_{thr}$)}
In \Cref{eq:zthr}, the threshold $\mathcal{Z}_{thr}$ determines which documents to filter based on \gls{au}, to remove approximately 5-10\% of documents that are poorly aligned with the target domain.
Since our density-based filtering relies on normalized scores, the threshold can be adjusted to control the removal percentage.
We evaluated both retrieval performance and document filtering ratios for $\mathcal{Z}_{thr} \in \{1.0, 1.5, 2.0, 2.5, 3.0\}$.
As presented in \Cref{tab:hyperparam_sensitivity}, using coCondenser, $\mathcal{Z}_{thr}=1.5$ achieves optimal performance of 29.27 nDCG@10 while filtering approximately 5\% of documents.
More strict thresholds ($\mathcal{Z}_{thr}=1.0$) remove an excessive proportion of documents (10\%), resulting in degraded performance of 28.57 nDCG@10, whereas more loosened thresholds ($\mathcal{Z}_{thr} \geq 2.0$) retain noisy outliers, leading to performance degradation (nDCG@10 $\leq$ 28.82).
% We further verified that with $\mathcal{Z}_{thr}=1.5$, the proportion of 
% aleatoric-uncertain samples falls within the expected 5–15\% range across other 
% datasets: trec-covid (9.32\%), robust04 (13.95\%), quora (5.73\%), trec-news 
% (15\%), and hotpotqa (5\%).

\paragraph{Balance Weight ($\lambda$)}
We set $\lambda=0.5$ to equally balance uncertainty and diversity in \Cref{eq:score-balance}.
To validate this choice, we conducted a hyperparameter sweep using DPR with $\lambda \in \{0.1, 0.3, 0.5, 0.7, 0.9\}$.
We excluded the boundary values of 0 and 1, as these would entirely neglect one component.
As demonstrated in \Cref{tab:hyperparam_sensitivity}, $\lambda=0.5$ yields optimal performance, substantially outperforming extreme values that disproportionately emphasize either uncertainty ($\lambda=0.1$, nDCG@10=15.37) or diversity 
($\lambda=0.9$, nDCG@10=15.33).
The balanced configuration of $\lambda=0.5$ confirms that neither metric alone is sufficient for effective sampling.

\paragraph{Smoothing Factor ($\alpha$)}
For early stopping criteria, we applied Exponential Moving Average (EMA) smoothing factor with $\alpha=0.4$ to address noticeable fluctuations in the uncertainty measure.
With a step size of 10, $\alpha=0.4$ effectively corresponds to a weighted window of approximately 4 steps.
This parameter selection is critical for early stopping: certain datasets terminate after only 2.5k samples (5 steps), and excessive smoothing ($\alpha > 0.5$) over-attenuates the signal, pulling it too strongly toward the global average and obscuring the underlying trend.
We selected $\alpha=0.4$ to balance noise reduction and trend preservation given the limited number of steps.
As illustrated in \Cref{fig:abl:ema}, smaller values (e.g., $\alpha=0.3$) exhibit similar early stopping behavior and do not noticeably affect performance, but demonstrate reduced stability on average compared to $\alpha=0.4$.

\section{Conclusion}
% We studied document selection for \gls{uda} retrievers and found that DUQGen’s diversity-only strategy tends to oversample low-density and high-confidence regions, which we interpret through aleatoric (data) and epistemic (model) uncertainty. 
% Building on this view, \ours approach jointly accounts for both uncertainties by filtering noisy, unreliable documents and emphasizing the informative documents to the model, while maintaining diversity across clusters. 
% This model-conditioned perspective contrasts with DUQGen’s model-agnostic diversity and yields consistent gains across five large BEIR corpora and three retrievers, using fewer pseudo-queries and enabling more sample-efficient training.
% These results highlight uncertainty-aware sampling as a robust direction for budget-efficient domain adaptation, and we see promising extensions toward learned uncertainty proxies and adaptive budget scheduling.

We studied document selection for \gls{uda} retrievers and found that DUQGen's diversity-only strategy tends to oversample low-density and high-confidence regions.
\ours addresses this by filtering noisy documents via \gls{au} and prioritizing informative ones via iterative \gls{eu} sampling.
% The iterative sampling-training loop tackled the performance convergence that can be raised due to \gls{eu} shifts, enabling gradual improvement by continuous sampling on underexplored high \gls{eu} regions.
This model-conditioned approach yields consistent gains across five BEIR corpora and four retrievers with fewer pseudo-queries.
Our results highlight uncertainty-aware sampling as a promising direction for budget-efficient domain adaptation.

\section{Limitations}
While our method consistently improves retrieval performance for single-vector retrievers, our treatment of multi-vector and re-ranking architectures is preliminary. As reported in \Cref{sec:appendix:other-results}, we approximate single-vector behavior by applying pooling and a shared vocabulary projection layer. However, these processes depart from the model's native training objectives, limiting interpretability. Developing protocols that align with each architecture's objective is a direction for future work. Finally, our estimate of the target distribution relies on IDF statistics. Exploring richer distribution measures (e.g., topic-conditioned statistics, document-frequency variants) is another promising direction.

Lastly, our resampling penalty mitigates bias toward dominant clusters but does not explicitly account for rare minority topics that might be relevant to the target domain. 
In domains with highly skewed topic distributions, such topics may still be underrepresented in the final training set, potentially limiting adaptation on those topics. 
Evaluating the impact of our sampling strategy on minority topic coverage is an important direction for future work.

% Our method primarily focus on the document selection problem. Another direction explores query generation methods. In future work, we plan to extend ours to document-query joint utilities such that jointly optimizes the document selection with pseudo query generation.
% Our method consistently improves retrieval performance across all retrievers, but the margin of gain diminishes as model capacity grows. 
% Similar trends have been observed in document and query expansion research \cite{weller2023generative}, suggesting that high-capacity models inherently yield smaller adaptation gains. 
% In future work, we plan to investigate the underlying causes of this phenomenon and, based on those insights, develop techniques to strengthen domain adaptation in larger, more capable retrievers.

\section*{Acknowledgments}
% BRL
This work was supported by the National Research Foundation of Korea(NRF) grant funded by the Korea government(MSIT) (No. RS-2024-00414981), 
% 상식추론
Institute of Information \& communications Technology Planning \& Evaluation (IITP) grant funded by the Korea government (MSIT) (No. 2022-0-00077/RS-2022-II220077, AI Technology Development for Commonsense Extraction, Reasoning, and Inference from Heterogeneous Data),
% GSP
and Institute of Information \& communications Technology Planning \& Evaluation (IITP) grant funded by the Korea government(MSIT) [NO.RS-2021-II211343, Artificial Intelligence Graduate School Program (Seoul National University)].


\bibliography{anthology, Z_bibliograpy/domain_adaptation_classical, Z_bibliograpy/others}

\newpage
\appendix

\section{Appendix}

\subsection{Algorithmic Description}
\label{sec:appendix:algo}
Algorithm \ref{alg:unite} describes the formal procedure of \ours. 
Initially, the corpus is refined by filtering out high \gls{au} samples (Line 1). 
The method then proceeds iteratively through $T$ rounds. 
In each iteration, we first assess the \gls{eu} of the remaining candidates and monitor convergence (Lines 4–7). 
We then perform adaptive sampling, where cluster weights are dynamically adjusted to penalize redundancy, ensuring the selection of a diverse batch $S_t$ (Lines 9–13). 
Finally, the retriever $\theta$ is updated using pseudo-queries generated from $S_t$ (Lines 15–16). 
This iterative interaction allows the model to progressively adapt to the target domain by focusing on informative yet reliable samples.
\begin{algorithm*}[t]
\caption{UnIte: Uncertainty-based Iterative Document Sampling}
\label{alg:unite}
\begin{algorithmic}[1]
\Require Unlabeled Corpus $\mathcal{C}$, Initial Retriever $\theta_0$, Iteration limit $T$, Batch size $B$
\State $\mathcal{C}' \leftarrow \text{AU\_Filtering}(\mathcal{C})$ \Comment{Remove high aleatoric uncertainty outliers}
\State Initialize sampled set $\mathcal{P} \leftarrow \emptyset$
\For{$t = 1$ to $T$}
    \State \textbf{// Step 1: EU Estimation}
    \State Compute $U_k(d; \theta_{t-1})$ for all $d \in \mathcal{C}'$ using Eq. (3)
    \State Calculate domain average EU $\bar{U}^{(t)}$
    \If{$\text{Plateau}(\bar{U}^{(t)})$} \textbf{break} \EndIf \Comment{Early Stopping}
    
    \State \textbf{// Step 2: EU Sampling}
    \For{cluster $c_j$ in Clusters}
        \State Update weight $w_j$ based on prior selection $\mathcal{P}_j$ using Eq. (5) \Comment{Resampling Penalty}
    \EndFor
    \State $S_t \leftarrow \text{Select } B \text{ docs maximizing Eq. (4) with weights } w$
    \State $\mathcal{P} \leftarrow \mathcal{P} \cup S_t$
    
    \State \textbf{// Step 3: Model Update}
    \State Generate pseudo-queries for $S_t$
    \State Update retriever $\theta_t \leftarrow \text{Train}(\theta_{t-1}, S_t)$
\EndFor
\State \Return Updated Retriever $\theta_t$
\end{algorithmic}
\end{algorithm*}

\subsection{Fine-tuning details}
\label{sec:appendix:fine-tuning}
We form a synthetic training set $\{(q_d,d)\}$ for $d\in S$, mining negatives via:
\begin{itemize}
  \item \emph{In‐batch negatives:} other documents $d'\in S$ in the same batch.
  \item \emph{Contriever hard negatives:} bottom negatives of the top-100 Contriever retrievals on $\mathcal{D}\setminus\{d\}$.
\end{itemize}
Bi‐encoder models (DPR, coCondenser, COCO-DR) use an InfoNCE loss. DPR was fine-tuned with a batch size of 32, learning rate of $2e^{-5}$, AdamW optimizer with weight decay of $1e^{-2}$, eps of $1e^{-8}$, and maximum sequence length of 509. coCondenser was fine-tuned with a batch size of 32, learning rate of $5e^{-6}$, and other training details follow Tevatron \cite{Gao2022TevatronAE}. COCO-DR was fine-tuned with a batch size of 32, learning rate of $1e^{-6}$ , and also followed the Tevatron train setup. \footnote{Trained using RTX 3090 GPU with 24GB memory}
Fine‐tuning runs on NVIDIA 3090 GPU, with each epoch completing in approximately 10 minutes.
For training GPL Baseline, we follow the data generation process and only modified the selected number of documents to 5k. Learning rate is modified to match the same setup as mentioned above.

\subsection{AU Approximation}
\label{sec:appendix:au-approx}
\begin{figure*}[!t]
{
\centering
    \includegraphics[width=0.7\textwidth]{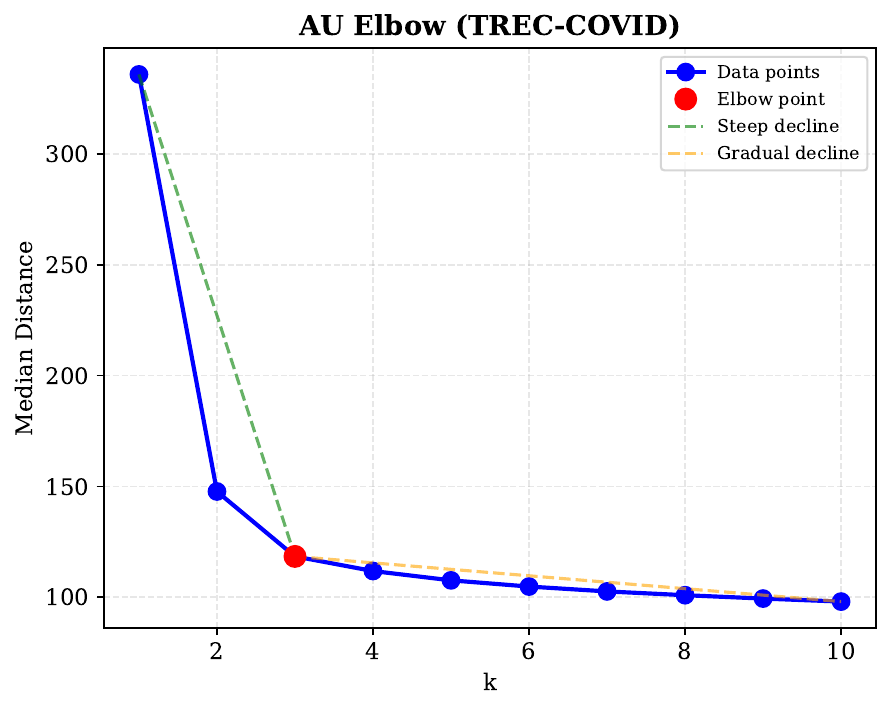}
    \caption{
        The medians of the lexical kNN distances across various k-values are illustrated.
    }
    \label{fig:au-elbow}
}
\end{figure*}
To find the optimal number of hyper-parameter $k$ in AU approximation using lexical k-NN distance, we use the elbow \cite{elbow_thorndike1953belongs} method in each domain. \Cref{fig:au-elbow} illustrates the elbow point detected in the TREC-COVID domain. For all 5 domains selected in \ours share the same elbow point as the k-value of 3.

% \subsection{Exponential Moving Average}
% \label{sec:appendix:ema}
\begin{figure*}[t] % 2단 논문의 한쪽 단(column)에 넣으려면 figure, 전체 폭을 쓰려면 figure*
  \centering

  % --- 첫 번째 줄 (Row 1) ---
  \begin{subfigure}[t]{0.48\linewidth} % 너비를 절반 정도로 설정
    \centering
    \includegraphics[width=\linewidth]{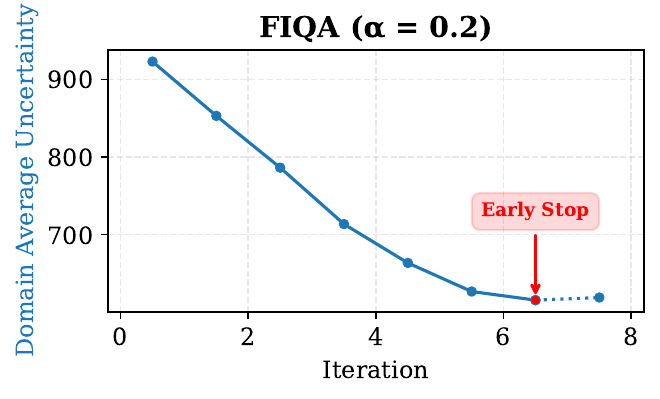} 
    % \caption{Description 1}
    \label{fig:1a}
  \end{subfigure}
  \hfill % 좌우 간격 자동 조절
  \begin{subfigure}[t]{0.48\linewidth}
    \centering
    \includegraphics[width=\linewidth]{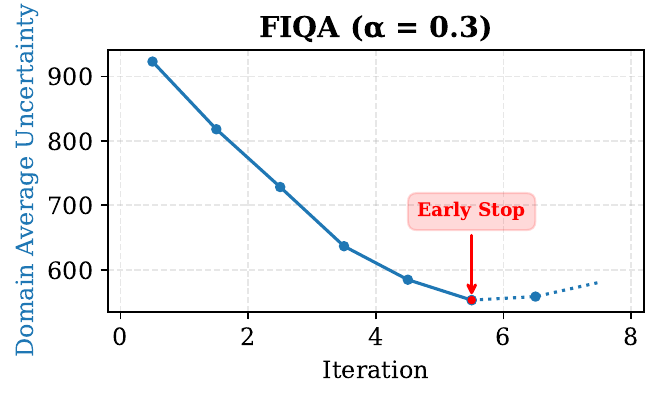}
    % \caption{Description 2}
    \label{fig:1b}
  \end{subfigure}
  
  \vspace{10pt} % 위아래 줄 간격 조절 (필요에 따라 숫자 변경)

  % --- 두 번째 줄 (Row 2) ---
  \begin{subfigure}[t]{0.48\linewidth}
    \centering
    \includegraphics[width=\linewidth]{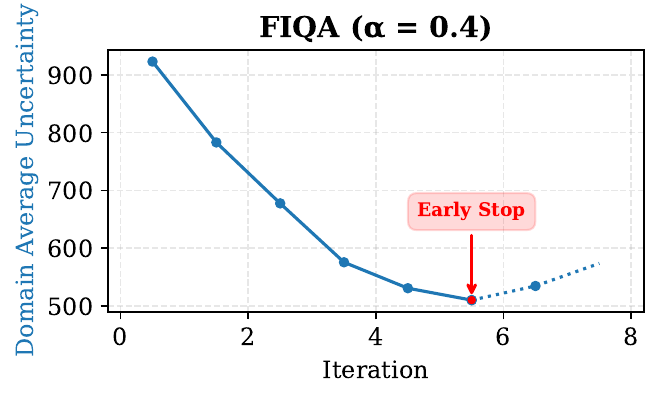}
    % \caption{Description 3}
    \label{fig:1c}
  \end{subfigure}
  \hfill
  \begin{subfigure}[t]{0.48\linewidth}
    \centering
    \includegraphics[width=\linewidth]{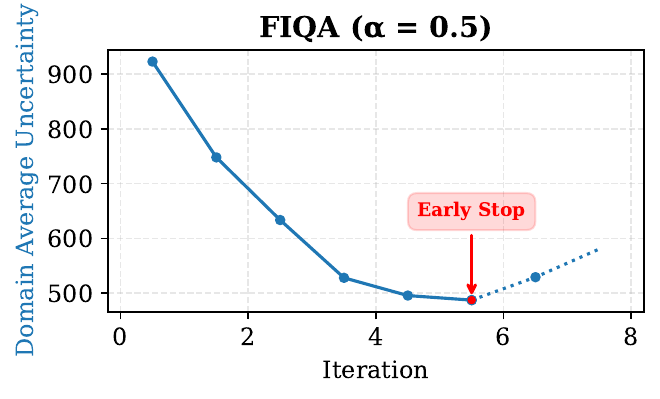}
    % \caption{Description 4}
    \label{fig:1d}
  \end{subfigure}

  \caption{
    Comparison of smoothed graph trends across different settings of smoothing factors.
  }\label{fig:abl:ema}
\end{figure*}
% We apply Exponential Moving Average (EMA) smoothing because the uncertainty measures exhibit noticeable fluctuations. When the step size is 10, setting $\alpha = 0.4$ effectively corresponds to a weighted window of about 4 steps.
% Due to early stopping, some datasets terminate after only about 2.5k samples (that is, 5 steps). In such
% cases, using $\alpha$  > 0.5 would oversmooth the curve, pulling it too strongly toward the global average and obscuring the underlying trend. 
% To balance smoothing and trend preservation under this limited number of steps, we choose $\alpha = 0.4$. 
% Smaller values, such as 0.3 show similar behavior on early stopping and do not noticeably affect performance, as you can find the impact on \Cref{fig:abl:ema}.

\subsection{Other models results}
\label{sec:appendix:other-results}
\begin{table*}[t]
  \centering
  \small
  \scalebox{0.95}{
  \renewcommand{\arraystretch}{1.2}
  \begin{tabular}{ 
      l 
      l 
      || c c c c c
      || c 
    }
    \toprule
    \multicolumn{1}{c}{
        \multirow{2}{*}{\textbf{Retriever}}
    }
      & \multicolumn{1}{c||}{
            \multirow{2}{*}{\shortstack[c]{\textbf{Adaptation}\\\textbf{Method}}}
      } 
      & \multicolumn{5}{c||}{\textbf{Large Corpus}} 
      & \multicolumn{1}{c}{
            \multirow{2}{*}{\textbf{Total AVG}}
      } \\
    \cmidrule(lr){3-7}
      & 
      & \textbf{TC} & \textbf{RB} & \textbf{QR} 
      & \textbf{TN} 
      & \textbf{HQ} \\ 
    \midrule
    \multicolumn{8}{l}{\textbf{First-stage Retriever}} \\
    \midrule
    BM25
      & —      
      & 65.59 & 40.70 & 78.9 
      & 39.8 & 60.3
      & 44.49 \\
    \midrule
    \multirow{3}{*}{ColBERT}  
      & — $\dag$
      & 70.6 & 39.2 & \underline{85.3}
      & \underline{39} & 59
      & 58.62 \\
      & DUQGen  
      & \textbf{74.18} & 44.95 & \textbf{85.57}
      & 36.93 & \textbf{63.44}
      & \textbf{61.01} \\
      & \ours   
      & \underline{73.43} & \textbf{46.37} & 85.09
      & \textbf{37.48} & \underline{60.64}
      & \underline{60.6} \\
    \midrule\midrule
    \multicolumn{8}{l}{\textbf{Reranking BM25 Top-100}} \\
    \midrule
    \multirow{4}{*}{monoT5} 
      & — $\dag$      
      & 81.39 & 51.81 & 84.65 & \underline{45.9} & 69.79 & 66.71 \\
      & InPars$\dagger$ 
      & 80.3 & 51 & — & — & — & - \\
      & DUQGen  
      & \underline{84.76} & \textbf{54.04} & \textbf{88.17} & 45.04 & \textbf{71.54} & \underline{68.71} \\
      & \ours   
      & \textbf{85.09} & \underline{53.91} & \underline{88.1} & \textbf{45.79} & \underline{71.25} & \textbf{68.83} \\

    \bottomrule
  \end{tabular}
}
  \caption{
  Retrieval performance (nDCG@10) on BEIR across late interaction and reranking model and adaptation methods. 
  \textbf{Bold} entries mark the highest performances per dataset for each retriever, while the \underline{Underlined} entries indicate the second-highest. 
  "AVG" columns report the overall average. 
  \dag indicates values taken from the original paper, and * denotes a statistically significant improvement over DUQGen (p<0.05). 
  }
  \label{tab:appendix-results}
\end{table*}

In \ours method, we assume that there is a single-vector embedding of each document, and apply the MLM Head of the base model to acquire the logit scores.
ColBERT is a multi-vector late-interaction model, and MonoT5 reranker is an Encoder-Decoder model. In both cases, we can't naturally derive single-vector embeddings. 
For the ColBERT model, we drop the token-level embedding projection layer and apply mean-pooling to get the single-vector embeddings. Also, for the MonoT5 model, we take the mean-pooling for the Encoder's last hidden state and use it as a single-vector embedding. For the language modeling head, T5 LM Head is used. Since this LM Head is not trained to accept the encoder hidden state, there might be some errors in the underlying logit value. The results are illustrated in \Cref{tab:appendix-results}. Although not every domain wins against the baseline, the average performance of \ours outperforms the baseline in MonoT5.

ColBERT uses a cross‐entropy over token interactions. We fine-tuned with official hyperparameters, which are the batch size of 32, the learning rate of $3e^{-6}$, and the maximum sequence length of 300.
MonoT5 uses a next-token prediction loss, and fine-tuned MonoT5-base using a batch size of 8, a learning rate of $2e^{-5}$, Adafactor optimizer with weight decay of $5e^{-5}$, and the warp-up ratio of 0.1, regression loss, and maximum sequence length of 512.

\subsection{Dataset Statistics}
\label{sec:appendix:dataset}
% \begin{table*}[t]
% \centering
% \scalebox{0.87}{
% \def\arraystretch{1.2}
% \begin{tabular}{l|c|c|c|c|c|c}
% \noalign{\hrule height 1pt}
% \multicolumn{1}{c|}{} & \textbf{SciFact} & \textbf{TREC-COVID} & \textbf{FiQA} & \textbf{BioASQ} & \textbf{Robust04} & \textbf{CQADupStack} \\ \hline
% Domain & Scientific & Bio-Medical & Financial & Bio-Medical & News & Forum \\
% Total \# Queries & 300 & 50 & 648 & 500 & 249 & 13,145 \\
% Total \# Documents & 5.2K & 129.2K & 57.6K & 1.0M & 528.2K & 457.2K \\
% Average Query Length & 12.4 & 10.6 & 10.8 & 8.1 & 15.3 & 8.6 \\
% Average Document Length & 213.6 & 210.3 & 132.2 & 204.1 & 466.4 & 129.1 \\
% \noalign{\hrule height 1pt}
% \end{tabular}
% }
% \caption{Statistics of the 6 subtasks of BEIR benchmark in our experiments. We follow the same preprocessing procedure of GPL.}
% \label{tab:dataset}
% \end{table*}

\begin{table*}[t]
\centering
\scalebox{0.9}{
\def\arraystretch{1.1}

%%%%%%%%%%%%%%%%%%%%%%%%%%%%%%%%%%%%%%%%%%%%%%%%%%%%%%%%%%%%%%%%%%%%%%%%%%%%%%%%%
% - TC
%%%%%%%%%%%%%%%%%%%%%%%%%%%%%%%%%%%%%%%%%%%%%%%%%%%%%%%%%%%%%%%%%%%%%%%%%%%%%%%%%
% \begin{tabular}{l|c|c|c|c|c}
% % \noalign{\hrule height 1pt}
% \hline
% \hline
% \multicolumn{1}{c|}{} & \textbf{SciFact} & \textbf{SciDocs} & \textbf{FiQA} & \textbf{NFCorpus} & \textbf{Robust04} \\ \hline
% Domain & Scientific & Scientific & Financial & Bio-Medical & News \\
% Total \# Queries & 300 & 1000 & 648 & 323 & 249 \\
% Total \# Documents & 5.2k & 25.7k & 57.6k & 3.6k & 528.2k  \\
% Average Query Length (words) & 12.4 & 9.4 & 10.8 & 3.3 & 15.3  \\
% Average Document Length (words) & 213.6 & 176.2 & 132.2 & 232.3 & 466.4 \\
% Relevant Document / Query & 1.1 & 4.9 & 2.6 & 38.2 & 69.9 \\

%%%%%%%%%%%%%%%%%%%%%%%%%%%%%%%%%%%%%%%%%%%%%%%%%%%%%%%%%%%%%%%%%%%%%%%%%%%%%%%%%
% + TC
%%%%%%%%%%%%%%%%%%%%%%%%%%%%%%%%%%%%%%%%%%%%%%%%%%%%%%%%%%%%%%%%%%%%%%%%%%%%%%%%%
\begin{tabular}{l|c|c|c|c|c}
% \noalign{\hrule height 1pt}
\hline
\hline
\multicolumn{1}{c|}{} & \textbf{TREC-COVID} & \textbf{Robust04} & \textbf{Quora} & \textbf{TREC-NEWS} & \textbf{HotpotQA} \\ \hline
Domain & Bio-Medical & News & Quora & News & Wikipedia \\
Total \# Queries & 50 & 249 & 10,000 & 57 & 7,405 \\
Total \# Documents & 171.3k & 528k & 522k & 594k & 5.2M \\
% Average Query Length (words) & 12.4 & 9.4 & 10.8 & 3.3 & 15.3 & 6.6 &  \\
% Average Document Length (words) & 213.6 & 176.2 & 132.2 & 232.3 & 466.4 & 160.8 \\
Relevant Document / Query & 1326.7 & 1250.6 & 1.6 & 53.35 & 2.1 \\
% \noalign{\hrule height 1pt}
\hline
\hline
\end{tabular}
}
\caption{Detailed statistics of the seven subtasks in the BEIR Benchmark that are employed in our experiments. 
This table includes the number of queries, the number of documents, and the number of relevant documents for a query, for each subtask. 
}
\label{tab:dataset}
\end{table*}
% \todo{update \autoref{tab:dataset} dataset}

% values are annotated here.
% https://docs.google.com/spreadsheets/d/1A7SeaYHtrFGCv-L7uWkWgOiSw5HWCegDkPorCg5l4Xk/edit?usp=sharing
In this paper, we focus on five BEIR benchmark datasets \cite{beir}:
TREC-COVID, Robust04, Quora, TREC-NEWS, and HotpotQA.
The statistics for each dataset, including the number of documents, test queries, and relevant documents per query, can be found in \Cref{tab:dataset}.

\subsection{Licenses for Artifacts}
All datasets and pretrained models were used strictly for research purposes, in accordance with their intended use and license terms. 
All datasets are English-only, and no demographic annotations were used. 
We complied with all license restrictions, including non-commercial clauses where applicable (e.g., DPR under CC BY-NC 4.0). 
The datasets are publicly available via BEIR on GitHub\footnote{\url{https://github.com/beir-cellar/beir}} and Hugging Face\footnote{\url{https://huggingface.co/BeIR}} under their respective licenses. 
Pretrained models were obtained from Hugging Face under their original licenses: DPR (CC BY-NC 4.0), coCondenser (Apache-2.0), COCO-DR (MIT), ColBERT (MIT), and MonoT5 (MPL-2.0). 
We will release our code and training scripts for research and educational use only, which is compatible with the access conditions of the original resources.

\subsection{Use of Packages}
We used \texttt{spaCy}~\cite{spacy_Honnibal_spaCy_Industrial-strength_Natural_2020} for text pre-processing before document sampling, including tokenization and stop-word removal for inverse document frequency (IDF) computation.
We employed \texttt{Tevatron}~\cite{Gao2022TevatronAE} to train \texttt{coCondenser} and \texttt{COCO\mbox{-}DR}.

\clearpage
\onecolumn
\subsection{Prompts}
\label{sec:appendix:prompt}

\begin{figure*}[!h]
\begin{tcolorbox}[]
\textbf{\textcolor{teal}{Example 1:}}

\textbf{\textcolor{cyan}{Document:}} December 25, 1990, Tuesday, Orange County EditionA mobile-home fire that killed an elderly woman Sunday night was accidental andstarted in her bed, Orange County Fire Department officials said Monday."Some sort of smoking materials in the bedding ignited the fire," said KathleenCha, a County Fire Department spokeswoman.The 75-year-old woman, whose name has been withheld pending notification ofrelatives, …

\textbf{\textcolor{red}{Relevant Query:}} What caused the fatal mobile-home fire in Dana Point that killed an elderly woman?.

\textbf{\textcolor{teal}{Example 2:}}

\textbf{\textcolor{cyan}{Document:}} 930818A LABOUR government would impose a levy of up to 1.5 per cent of payrollcosts on companies which failed to comply with training guidelines, MrGordon Brown, shadow chancellor, said yesterday.The levy, intended to help pay for upgrading government training programmes,compares with earlier plans for a maximum levy of 0.5 per cent on allcompanies not spending that amount.The revised proposal emerged in a paper for Labour's annual conference nextmonth, in which Mr Brown further distances the party from thehigher-taxation manifesto on which it fought the 1992 general election.Promising to cut taxes 'if I can', Mr Brown confirmed the Labourleadership's determination to discard the party's redistributionist image.'Labour is not against wealth, nor will we seek to penalise it,' he said.Mr Brown said the revised training proposals were aimed at encouragingcompanies to develop their own training programmes, rather than rely on thegovernment.'There are a large number of companies which are failing to make thetraining investment which is necessary. That is not only harming the countryas a whole, it is harming those companies which are prepared to make theinvestment because they are finding that their trained workers are being ...

\textbf{\textcolor{red}{Relevant Query:}} What are Labour's proposed training levy guidelines for companies and the rationale behind them?

\textbf{\textcolor{teal}{Example 3:}}

\textbf{\textcolor{cyan}{\textbf{Document:}}} BFN[Unattributed report: "A Loose EU Is Not Necessarily to OurAdvantage"][Text] As a member of the European Union [EU], Finlandmust not become subservient to the interests of any of the bigpowers in the EU. Finland must not align itself with either theBritish or the French ideology, but act as defender of theinterests of Europe's northeast corner. A loose EU is notnecessarily advantageous to Finland.These were the points that SDP [Social Democratic Party]Chairman Paavo Lipponen stressed in his speech at a seminarorganized by the Trade and Industry Delegation on Wednesday [4May]. Lipponen pointed out that Finland has eight months toprepare its membership policy and get ready to take fulladvantage of membership. He added that Finland cannot functionfor a single day in the EU without a clearly defined platform ofpolicy."Will it be the French or the British philosophy? The Frenchfocus on finality, a clearly defined goal for a closelyintegrated, federalist EU. The British prefer to advancepragmatically and favor a loose community until this is provenwrong." …

\textbf{\textcolor{red}{Relevant Query:}} What stance does Finland's SDP Chairman Paavo Lipponen take on a loose versus closely integrated EU, and how does this relate to Finland's interests?

\textbf{\textcolor{teal}{Example:}}

\textbf{\textcolor{cyan}{Document:}} \{document\_text\}

\textbf{\textcolor{red}{Relevant Query:}}

\end{tcolorbox}

\caption{Prompt template with in-context examples for query generation for the Robust04 Dataset.}
\label{fig:fewshot-prompt}
\end{figure*}

\clearpage
\twocolumn
\subsection{Use of AI Assistants}
We used ChatGPT for grammatical corrections.

%%%%%%%%%%%%% CUSTOM %%%%%%%%%%%%%
% \clearpage
% \section{Custom Area for draft writing}
% \tableofcontents
% \clearpage
% \input{A_sections/__outline}

% \clearpage
% \listoftodos[List of suggested changes]

% \clearpage

%%%%%%%%%%%%%%%%%%%%%%%%%%%%%%%%%%
\end{document}